\documentclass[twocolumn,prx,twocolumn,superscriptaddress]{revtex4}
\pdfoutput=1 
\usepackage[latin9]{inputenc}
\usepackage{color}
\usepackage{amsmath}
\usepackage{amssymb}
\usepackage{graphicx}
\usepackage{tikz}
\usepackage{verbatim}
\usepackage{amsmath}
\usepackage{amssymb}
\usepackage{graphicx}
\usepackage{tikz}
\usepackage{amsmath,bm,amssymb,latexsym,galois,amsthm,euscript,dsfont}
\usepackage{amsmath,bm,amssymb,latexsym,galois,amsthm,euscript}
\usepackage[all,cmtip,knot]{xy} 
\usepackage{mathrsfs}
\usepackage[unicode=true,bookmarks=true,bookmarksnumbered=false,bookmarksopen=false,breaklinks=false,pdfborder={0 0 1},backref=false,colorlinks=true]{hyperref}
\usepackage{algorithm}
\usepackage{algorithmic}

\hypersetup{
    colorlinks,
    linkcolor={red!50!black},
    citecolor={blue!90!black},
    urlcolor={blue!90!black}
}

\makeatletter

\usepackage{amsfonts}\usepackage{tabularx}\usepackage{dcolumn}\usepackage{bm}\usepackage{graphicx}\usepackage{epstopdf}

\newcommand{\orcid}[1]{\href{https://orcid.org/#1}{\includegraphics[width=8pt]{orcid}}}

\hypersetup{urlcolor=blue}

\theoremstyle{definition}

\theoremstyle{remark}

\makeatother

\begin{document}
	
	\title{The Exclusivity Principle Determines the Correlation Monogamy}
	\author{Zhih-Ahn Jia}
	\email{giannjia@foxmail.com}
	\affiliation{Key Laboratory of Quantum Information, Chinese Academy of Sciences, School of Physics, University of Science and Technology of China, Hefei, Anhui, 230026, P. R. China}
	\affiliation{Synergetic Innovation Center of Quantum Information and Quantum Physics, University of Science and Technology of China, Hefei, Anhui, 230026, P. R. China}
	\author{Gao-Di Cai}
	\affiliation{Department of Physics, Shandong University, Jinan 250014, China}
	\author{Yu-Chun Wu}
	\email{wuyuchun@ustc.edu.cn}
	\affiliation{Key Laboratory of Quantum Information, Chinese Academy of Sciences, School of Physics, University of Science and Technology of China, Hefei, Anhui, 230026, P. R. China}
	\affiliation{Synergetic Innovation Center of Quantum Information and Quantum Physics, University of Science and Technology of China, Hefei, Anhui, 230026, P. R. China}
	\author{Guang-Can Guo}\affiliation{Key Laboratory of Quantum Information, Chinese Academy of Sciences, School of Physics, University of Science and Technology of China, Hefei, Anhui, 230026, P. R. China}
	\affiliation{Synergetic Innovation Center of Quantum Information and Quantum Physics, University of Science and Technology of China, Hefei, Anhui, 230026, P. R. China}
	\author{Ad\'{a}n Cabello}
	\email{adan@us.es}
	\affiliation{Departamento de F\'{\i}sica Aplicada II, Universidad de Sevilla, E-41012 Sevilla, Spain}

	\date{\today}
	
	\begin{abstract}
		Adopting the graph-theoretic approach to the correlation experiments, we analyze the origin of monogamy and prove that it can be recognized as a consequence of the exclusivity principle(EP). We provide an operational criterion for monogamy: if the fractional packing number of the graph corresponding to the union of event sets of several physical experiments does not exceed the sum of independence numbers of each experiment graph, then these experiments are monogamous. As applications of this observation, several examples are provided, including the monogamy for experiments of Clauser-Horne-Shimony-Holt (CHSH) type, Klyachko-Can-Binicio\u{g}lu-Shumovsky (KCBS) type, and for the first time, we give some monogamy relations of Svetlichny's genuine nonlocality. We also give the necessary and sufficient conditions for several experiments to be monogamous: several experiments are monogamous if and only if the Lov\'{a}sz number of the union exclusive graph is less than or equal to the sum of independence numbers of each exclusive graph.
	\end{abstract}

	\pacs{03.65.-w, 03.65.Ud}
	\maketitle
	
	\emph{Introduction.}\textemdash One of the characteristic features of classical correlation is its shareability among many parties, but the situation for quantum correlation is very different, it cannot be shared freely. The limitation on the shareability of quantum correlation is now known as monogamy relation, see, for example, Refs.\cite{Streltsov2012are,Koashi2004monogamy,dhar2016monogamy}. Since it was first qualitatively formulated by Coffman, Kundu and Wootters for quantum entanglement correlations using concurrence\cite{CKW2000}, monogamy as a quantum phenomenon has been studied in many types of correlations, such as Einstein-Podolsky-Rosen (EPR) steering\cite{Reid2013monogamy}, Bell nonlocality\cite{Pawlowski2009monogamy,Brunner2014bell,Jia2016}, contextuality\cite{Ramanathan2012generalized,Jia2016}, and contextuality-nonlocality\cite{Kurzynski2014fundamental,Jia2016}. Some experimental verifications for these monogamy relations are also reported\cite{Zhan2016realization}. From a practical perspective, monogamy has widespread applications in many areas of physics, including the derivation of security of quantum key distribution\cite{Pawlowski2010security}, determination of quantum critical point\cite{qin2016quantum}, giving a criterion of the maximally symmetry-breaking quantum ground states\cite{cianciaruso2016classical}, diagnosing topological edge states\cite{Meichanetzidis2016diagnosing}, even in the arguments of firewall problem of black holes\cite{susskind2013black,Lloyd2014}. However, the physical origin of these different kinds of monogamy is not yet very clear and there is no operational criterion for several experiments to be monogamous.

	In the typical correlation test scenario, some physicists run an experiment, in which a set of physical events $\mathcal{E}=\{e_i\}$ occur with the respective probabilities $\mathcal{C}(\mathcal{E})=\{p(e_i)\}$. Hereinafter, a physical event which we denote as $a_1,\cdots,a_n|x_1,\cdots,x_n$ is obtaining $a_1,\cdots,a_n$ upon measuring $x_1,\cdots,x_n$ on a physical system, assume that the preparation of the system is reproducible, then by repeating the experiment many times, we can collect the probabilities of $p(a_1,\cdots,a_n|x_1,\cdots,x_n)$ for each event $a_1,\cdots,a_n|x_1,\cdots,x_n$. To check if the observed statistics $\mathcal{C}(\mathcal{E})$, which we refer to as the experimental correlation or physical behavior as in some literature\cite{Brunner2014bell}, satisfies some physical principle $P$ (or some theoretical model $\mathcal{T}=\{P_1,\cdots,P_n\}$ which is nothing more than an assemblage of physical principles), we can calculate a test parameter $\mathcal{I}_P$, which is a (not necessarily) real-valued bounded function with variables being all involved probabilities, and find out if the resulting value lies in the range allowed by the principle $P$.

	In many cases, the $P$-character range $\mathcal{R}_P\subseteq \mathbb{R}$ is just an interval. For instance, in Bell's local hidden variable (LHV)\cite{bell1964} test scenario, if test parameter is chosen as Clauser-Horne-Shimony-Holt (CHSH) parameter\cite{CHSH}, the LHV-range is $\mathcal{R}_{LHV}=[-2,2]$; in Kochen-Specker's noncontextual hidden variable (NCHV) test scenario\cite{kochen1967problem}, if the test parameter is chosen as Klyachko-Can-Binicio\u{g}lu-Shumovsky (KCBS) parameter \cite{Klyachko2008simple}, its NCHV-range is $\mathcal{R}_{NCHV}=[-3,5]$. The ranges allowed by quantum mechanics of CHSH and KCBS parameters are $[-2\sqrt{2},2\sqrt{2}]\supseteq \mathcal{R}_{LHV}$ and $[5-4\sqrt{5},5]\supseteq \mathcal{R}_{NCHV}$ respectively, which indicates that quantum mechanics is beyond LHV model and NCHV model, but the violated values stop at a bound which is now generally referred to as Tsirelson's bound or quantum bound\cite{Tsirelson1980quantum}. Many efforts have been made to answer the question of what is the physical principle that prevents quantum mechanics from having a larger violation than one of quantum mechanics. Information causality\cite{pawlowski2009information}, local orthogonal principle\cite{fritz2013local}, measurement sharpness\cite{chiribella2014measurement}, and exclusivity principle (EP)\cite{Cabello2013simple,Yan2013} provide us with some rationales for why limits on quantum mechanics may exist. Among all of them, more and more evidences suggest that EP is suitable as a fundamental assumption of quantum mechanics. In this letter, we concern with a very relevant but somewhat weaker problem: why do different types of monogamy occur in a given theoretical framework, and how can we determine if several experiments are monogamous or not?

	Suppose we are running two experiments $\mathcal{E}=\{e_i|p(e_i)\}$ and $\mathcal{U}=\{u_i|p(u_i)\}$ simultaneously in a theoretical framework $\mathcal{T}$, which means that statistics $p(e_i)$ and $p(u_i)$ are obtained under the constraint of theory $\mathcal{T}$, e.g., in quantum mechanics it is $p(e_i)=\mathrm{tr}(\pi_{e_i}\rho)$ with $\pi_{e_i}$ the measurement corresponding to event $e_i$ and $\rho$ the prepared state and analogously for $p(u_i)$. In order to check if correlations $\mathcal{C}(\mathcal{E})$ and $\mathcal{C}(\mathcal{U})$ satisfy principle $P$ or not, we need to check that if $\mathcal{I}_1(p(e_i))\in [r_P^1,R_P^{1}]$ and $\mathcal{I}_2(p(u_i))\in [r_P^{2},{R}_P^{2}]$ or not, where $r_P^i$($R_P^i$) is the low (upper) bound allowed by $P$-principle. Without losing generality, hereinafter we assume that each test parameter $\mathcal{I}$ is positive since $\mathcal{I}$ is a bounded function with a lower bound $l$ (this bound is calculated only under the constraints imposed by probability theory, \emph{viz.}, Kolmogorov axioms), we can replace $\mathcal{I}$ with $\mathcal{I}'=\mathcal{I}-l$, the replacement do not influence the checking result. Two experiments are no-$P$ monogamous in $\mathcal{T}$ theory if
	\begin{equation}\label{monogamy}
		\mathcal{M}(\mathcal{I}_1,\mathcal{I}_2)\overset{\mathcal{T}}{\in}[m_P,M_P]\subseteq [\mathcal{M}(r_P^1,r_P^2),\mathcal{M}(R_P^1,R_P^2)]
	\end{equation}
	where $\mathcal{M}(\mathcal{I}_1,\mathcal{I}_2)$ is a monotonically increasing function which we refer to as the monogamy function and is often chosen as $\mathcal{I}_1+\mathcal{I}_2$, ``$\overset{\mathcal{T}}{\in}$'' indicates the tight bound allowed by $\mathcal{T}$ theory, and the respective lower (upper) bound $m_P$($M_P$) is referred to as monogamy score. The monogamy relation (\ref{monogamy})  means that if the experimental correlation $\mathcal{C}(\mathcal{E})$ is a no-$P$ correlation, \emph{viz.}, $\mathcal{I}_1(p(e_i))\not \in [r_P^1,R_P^1]$, then the correlation $\mathcal{C}(\mathcal{U})$ must be $P$-correlation, \emph{viz.}, the value of $\mathcal{I}_2$ must lie in the $P$-character interval $[r_P^2,R_P^2]$. We take the monogamy of nonlocality as an example, if Alice implements two  CHSH-type Bell experiments $\mathcal{E}_{AB}$ and $\mathcal{E}_{AC}$ with Bob and Charlie simultaneously, then the monogamy relation reads $\mathcal{I}_{AB}^{CHSH}+\mathcal{I}_{AC}^{CHSH}\in [-4,4]$, \emph{viz}, if Alice and Bob observe the violation of  $\mathcal{I}_{AB}^{CHSH}\in [-2,2]$ then Alice and Charlie must not observe the violation and vice versa.
	
	Since quantum mechanics is also a theoretical model consisting of a set of physical principles (actually, we haven't yet found out what these principles are), to explain the origin of no-$P$ monogamy in quantum mechanics, we must explain what constitutes principle of quantum mechanics can be used to derive the monogamy relation. There are some trials to explain the origin of different kinds of monogamy, for instance, the principle of no-disturbance\cite{Ramanathan2012generalized,Jia2016}(or more restricted no-signaling principle in Bell's scenario\cite{Pawlowski2009monogamy,Jia2016}) can be used to derive contextuality monogamy, correlation complementarity can be used to derive nonlocality monogamy\cite{Kurzynski2011correlation} and the Lorentz invariance in Bloch representation can be used to derive entanglement monogamy\cite{Eltschka2015monogamy}. But all these attempts provide only partial answers, and in many cases, their derivation can not give the tight monogamy bound in quantum mechanics.
	
	In this letter, we analyze the quantum correlations in detail and we find many important monogamy relations (monogamy of quantum contextuality, nonlocality, genuine nonlocality, and contextuality-nonlocality, etc.) appearing in quantum theory can be derived from EP. We prove that the monogamy derived from EP is tight, \emph{viz.}, the monogamy bound restricted by EP meets the bound restricted by quantum mechanics. We give the necessary and sufficient conditions for some experiments to be monogamous and as an application, we provide an operational criterion to determine if several experiments are monogamous. This sheds new light on the relationship between EP and quantum mechanics.
	
	We begin with a brief review of the EP.

	\emph{Exclusivity as a physical principle.}\textemdash Suppose that we are running an experiment $\mathcal{E}=\{e_i\}$, the test parameter for a physical property may have many types of formulations: it can be a sum $\sum_iw_ip(e_i)$ where each weight $w_i$ is a real number, like in Bell inequality, noncontextuality inequality; or some other function such as Shannon entropy, Tsallis entropy, R\'{e}nyi entropy and so on. Here, we focus on the sum-type test parameter, which is also the most studied form since Bell's seminal work\cite{bell1964}. A standard correlation test inequality of the experiment $\mathcal{E}$ is of the form:
	\begin{equation}\label{}
		\mathcal{I}_{\mathcal{E}}=\sum_iw(e_i)p(e_i)\overset{C}{\leq} R_C \overset{Q}{\leq} R_Q \overset{SQ}{\leq} R_{SQ},
	\end{equation}
	in which $R_C$, $R_Q$, and $R_{SQ}$ represent the classical bound, Tsirelson's bound, and supraquantum bound respectively. We assume each $w(e_i)$ to be a positive real number, this can be done by substituting probabilities of events that have a negative $w_i$ by unity minus the probability of the opposite events. Hereinafter, for simplicity, we will assume that all $w(e_i)$ is equal to 1, and all results can be applied to the general case by simply substituting all graph terms with the weighted graph terms. To explain why quantum mechanical violation of the inequality stops at the Tsirelson's bound, Cabello\cite{Cabello2013simple} and Yan\cite{Yan2013} suggest the EP which can be summarized as:
	
	\textsf{Exclusivity principle(EP)}: the sum of probabilities of pairwise exclusive events cannot exceed 1.
	
	Note that EP can be applied to a given target experiment without considering other experiments that may be carried out in another part of the universe, we refer to this kind of application of EP as weak EP, relatively, it can also be applied to extended experiments that shed some light on the correlations of target experiment, we refer to this case as strong EP. Tsirelson's bounds of  CHSH inequality\cite{CHSH} and KCBS inequality\cite{Klyachko2008simple} can be derived from EP\cite{Cabello2013simple,Yan2013}, more exactly, it should be strong EP.
	
	In Ref.\cite{Cabello2014graph}, Cabello, Severini, and Winter provide a graph theoretical approach to the correlation test experiment $\mathcal{E}$, it can be sketched as:
	\begin{displaymath}
		\xymatrix{
			\mathrm{inequality}~\mathcal{I_{\mathcal{E}}:}\ar@{<-->}[d]_{\mathrm{experiment}~\mathcal{E}} \\
			\mathrm{graph}~G_{\mathcal{E}}:}
		\xymatrix{
			R_{C} \ar@{<-->}[d]_{}  \\
			\alpha(G_{\mathcal{E}})}
		\xymatrix{
			\leq \\
			\leq}
		\xymatrix{
			R_{Q} \ar@{<-->}[d]_{} \\
			\vartheta(G_{\mathcal{E}})}
		\xymatrix{
			\leq \\
			\leq}
		\xymatrix{
			R_{SQ} \ar@{<-->}[d]_{}  \\
			\alpha^*(G_{\mathcal{E}})}
	\end{displaymath}
	where we associate an events graph $G_{\mathcal{E}}$ to each correlation experiment $\mathcal{E}$, in which a vertex represents an event $e_i$ and an edge represents an exclusive pair $(e_i,e_j)$. The correlation test parameter can be considered as a linear function $\mathcal{I}_{\mathcal{E}}$ of probabilities of related events. The classical bound $R_{C}$ is given by the independence number $\alpha(G)$ of the exclusive graph $G_{\mathcal{E}}$, which is the cardinality of the largest independent vertex set. The Tsirelson's bound is given by the Lov\'{a}sz number $\vartheta(G_{\mathcal{E}})$ of the exclusive graph, which is defined as $\vartheta(G_{\mathcal{E}})=\max\sum_{e_i\in V(G_{\mathcal{E}})}|\langle \phi|v_i \rangle|^2$ where the maximum is taken over all orthonormal representations of the complete graph of $G_{\mathcal{E}}$. The bound given by EP is then the fractional packing number $\alpha^*(G_{\mathcal{E}})$ of the exclusivity graph. See supplementary material \cite{supp} for the involved graph theoretical terminologies in more detail.
	
\begin{figure}
		\includegraphics[scale=0.9]{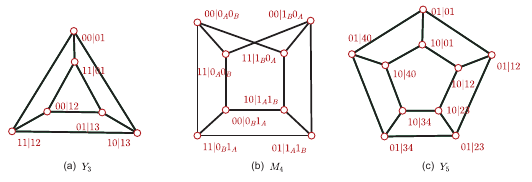}
		\caption{\label{fig:cycle} (color online). The depiction of the exclusive graph of the $n=3,4,5$ cycle inequalities, the graph of 4-cycle CHSH inequality is an M\"{o}bius ladder $M_4$, and the graph of $5$-cycle KCBS inequality is a prism graph $Y_5$.}
	\end{figure}

	Here we take the  CHSH inequality
	\begin{equation}\label{}
		\sum_{a\oplus b=xy}\sum_{x,y=0,1} p(a,b|x,y) \overset{C}{\leq} 3 \overset{Q}{\leq} 2+\sqrt{2}\overset{E}{\leq} 4,
	\end{equation}
	and KCBS inequality
	\begin{equation}\label{}
		\sum_{a\oplus b=1}\sum_{i=1}^{5}p(a,b|i,i+1) \overset{C}{\leq} 4 \overset{Q}{\leq} 2\sqrt{5}\overset{E}{\leq} 5,
	\end{equation}
	as two important examples. As depicted in Fig \ref{fig:cycle}, the exclusive graph of CHSH and KCBS experiments are a $4$-M\"{o}bius ladder $M_4$ and $5$-prism graph $Y_5$, it is obvious that graph theoretical terms coincide with the inequality bounds \cite{supp}. Note as a consequence of symmetry of prism graph $Y_5$, this KCBS inequality can be simplified into a five-vertex pentagon graph inequality
	\begin{equation}\label{}
		\sum_{i=1}^{5}p(0,1|i,i+1) \overset{C}{\leq} 2 \overset{Q}{\leq} \sqrt{5}\overset{E}{\leq} 5/2.
	\end{equation}
	
	These two inequalities can be extended to a unified $n$-cycle inequality\cite{liang2011specker,Araujo2013n-cycle} as
	\begin{eqnarray}\label{n-cycle-inequality}
		\mathcal{I}_n \overset{C}{\leq}  n-1 \overset{Q}{\leq} & \left\{
		\begin{array}{ll}
			\frac{2n\cos(\pi/n)}{\cos(\pi/n)+1},& n\in 2\mathbb{N}+1\\
			\frac{n(\cos(\pi/n)+1)}{2},& n\in 2\mathbb{N}
		\end{array}\right.
		\overset{E}{\leq} n,
	\end{eqnarray}
	where $\mathcal{I}_n$ is of the form $\sum_{a\oplus b=0}\sum_{\begin{subarray}{lcl}
			i=1,\cdots,n\\i\neq j\end{subarray}} p(ab|ii+1)+\sum_{a\oplus b\neq 0}p(ab|jj+1)$ or of its complementary form $\sum_{a\oplus b=1}\sum_{\begin{subarray}{lcl}
			i=1,\cdots,n\\i\neq j\end{subarray}} p(ab|ii+1)+\sum_{a\oplus b=0}p(ab|jj+1)$ for some $j\in\{1,\cdots,n\}$, and we make the convention that $n+1=1$.
	The graph of odd $n$-cycle inequality is a prism graph $Y_n$ which is isomorphic to the $2n$-vertex $(2,n)$ circulant graph $Ci_{2n}(2,n))$, the graph of even $n$-cycle inequality is a M\"{o}bius ladder graph $M_n$ which is isomorphic to the $2n$-vertex $(1,n)$ circulant  graph $Ci_{2n}(1,n))$, their independence numbers, Lov\'{a}sz numbers and frational numbers coincide with each physical correspondence.
	
	\emph{Origin of monogamy.}\textemdash An interesting aspect of these correlation inequalities is that the violations are monogamous, \emph{viz.}, violation of one inequality may lead to a unviolated value of the other inequality. It is a fundamental problem to pinpoint what part of quantum mechanics is responsible for these constraints. Here we proffer an explanation: monogamy is a consequence of EP.
	
	Consider two laboratories are running two testable correlation experiments with respective exclusivity sets $\mathcal{E}_i (i=1,2)$, where the word \emph{testable} means that the graph of the exclusivity set $G_{\mathcal{E}_i}$ has a distinct independence number, Lov\'{a}sz number and fractional packing number, and each graph theoretical term coincides with the physical bound of corresponding testing parameter $\mathcal{I}_{\mathcal{E}_i}$, this is equivalent to say that each exclusivity graph $G_{\mathcal{E}_i}$ contains, as induced subgraphs, odd cycles on five or more vertices and/or their complements\cite{Cabello2013basic}. If two experiments are implemented simultaneously, they can be regarded as an integral experiment $\mathcal{E}=\mathcal{E}_1 \sqcup \mathcal{E}_1$. Then there will be some exclusivity pairs $(e_i, u_j)$ with $e_i\in \mathcal{E}_1$ and $u_j\in \mathcal{E}_2$, which do not appear if we regard them as two independent experiments (\emph{viz.}, if they are not implemented simultaneously).
	
\begin{figure}
		\includegraphics[scale=0.65]{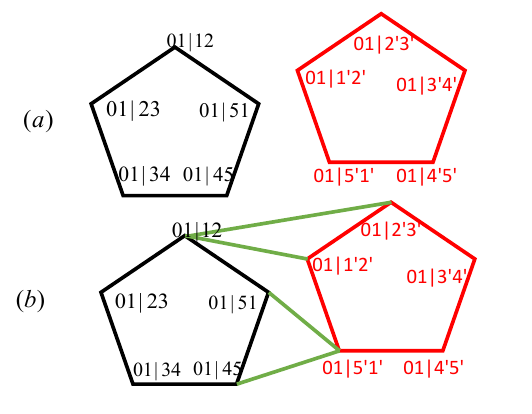}
		\caption{\label{fig:kcbsmonogamy} (color online). The depiction of the monogamy relation of two KCBS experiments, where the red and black pentagons are two exclusive graphs. (a) two experiments are implemented completely independent, the union graph is just two pentagons; (b) two experiments are implemented simultaneously, the union event set $\mathcal{E}=\mathcal{E}_{KCBS}\sqcup\mathcal{E}_{KCBS'}$ remain unchanged, but there will be some new exclusive pairs which are green edges in the exclusive graph of $\mathcal{E}$.}
	\end{figure}

	As depicted in Fig. \ref{fig:kcbsmonogamy}, if two non-contextuality experiments $\mathcal{E}_{KCBS}=\{01|12,01|23,01|34,01|45,01|51\}$ and $\mathcal{E}_{KCBS'}=\{01|1'2',01|2'3',01|3'4',01|4'5',01|5'1'\}$ (where we assume that the triple $1,1',2'$ are exclusive, \emph{viz.}, they can not get $0$ outcomes simultaneously, as is also the triple $4,5,5'$.) are not implemented simultaneously, their corresponding graph is just two pentagons, the red and black one as in Fig. \ref{fig:kcbsmonogamy}(a), but if they are implemented simultaneously, there will be some new exclusive pairs (four green edges as in Fig. \ref{fig:kcbsmonogamy}) like  $(01|12,01|1'2')$ and so on which does not exist before. Thus we will get the graph corresponding to $\mathcal{E}_{KCBS}\sqcup \mathcal{E}_{KCBS'}$ just as the whole graph in Fig. \ref{fig:kcbsmonogamy}(b). It is these new exclusive pairs that are responsible for monogamy relations.
	
	We are now in position to explain what is monogamy, and in what sense can a physical principle be regarded as the origin of it. Suppose that we are concerned about a physical principle $P$, like no-disturbance, no-signaling, and exclusivity. The principle will give some constraints while calculating the test parameter $\mathcal{I}_{\mathcal{E}_i}$, the maximal value of the $\mathcal{I}_{\mathcal{E}_i}$ restricted by the principle $P$ is called the $P$-bound, mathematically, $\mathcal{I}_{\mathcal{E}_i}\overset{P}{\leq} R_P^i$. If $P$ is chosen as classical mechanics, we have the classical bound $R_C$, similarly, for quantum mechanics, we have Tsirelson bound $R_Q$. We call two no-$P$ correlations monogamous in quantum mechanics if
	\begin{equation}\label{}
		\mathcal{I}_{\mathcal{E}_1}+\mathcal{I}_{\mathcal{E}_2}\overset{Q}{\leq} m_{P}(\mathcal{E})\leq R_P^1+R_P^2,
	\end{equation}
	where the maximal value of two simultaneous experiments allowed by quantum mechanics is called $P$-monogamy score in quantum mechanics, denoted as $m_P(\mathcal{E})$. This means that in quantum mechanical framework, \emph{viz.}, in our universe,  if  the first $P$-test experiment get the violated value $\mathcal{I}_{\mathcal{E}_1}(p(e_i))\geq R_P$, then the second $P$-test inequality cannot be violated. A physical principle $X$ can be regard as the origin of no-$P$ correlation monogamy if the maximal value of $\mathcal{I}_{\mathcal{E}_1}+\mathcal{I}_{\mathcal{E}_2}$ allowed by the principle $X$ is less than or equal to the $P$-monogamy score, \emph{viz.}, $\mathcal{I}_{\mathcal{E}_1}+\mathcal{I}_{\mathcal{E}_2}\overset{X}{\leq} R_X = m_{P}(\mathcal{E})$ .

	The appearance of extra exclusive pair when we run several experiments simultaneously make the fractional packing number $\alpha^*(G_{\mathcal{E}})$ which corresponds to the physical bound restricted by EP decrease in compare to the sum of individual fractional packing numbers $\alpha^*(G_{\mathcal{E}_1})+\alpha^*(G_{\mathcal{E}_2})$, if there are enough exclusivity pair to make $\alpha^*(G_{\mathcal{E}})\leq R_{C}^1+R_{C}^2=\alpha(G_{\mathcal{E}_1})+\alpha(G_{\mathcal{E}_2})$, then we get the monogamy relation. So monogamy is quantitative relation to evaluating the exclusive degree of two experiments. The above explanation can be summarized as the following result:
	
	\emph{Theorem~1.} Given several disjoint experimental event sets $\mathcal{E}_1,\cdots, \mathcal{E}_n$ with exclusive graphs $G_{\mathcal{E}_1},\cdots,G_{\mathcal{E}_n}$ respectively, we can make them into an assemblage of events $\mathcal{E}=\sqcup_{i=1}^n\mathcal{E}_i$ with exclusive graph $G_{\mathcal{E}}$, if we have the relation
	\begin{equation}\label{}
		\alpha^*(G_{\mathcal{E}})\leq \sum_{i=1}^n \alpha(G_{\mathcal{E}_i}),
	\end{equation}
	then these experiments are monogamous.
	
	See supplemental material for detailed proof. Since EP only concerns with the exclusive relation between physical events, it can be used in any type of correlation test scenario. To illustrate how EP can be utilized to establish monogamy relations, we give some examples.
	
	\emph{Example~1.~Monogamy of nonlocality:} consider three Bell-CHSH correlation test experiments $\mathcal{I}_{AB}$, $\mathcal{I}_{BC}$ and $\mathcal{I}_{CA}$ among Alice, Bob, and Charlie, with the corresponding exclusive sets of each run of the experiment are:
	\begin{equation}\label{3loop}
		\begin{array}{ccccc}
			\mathcal{E}_{AB}= &\{00|0_A0_B, & 00|0_B1_A, & 01|1_A1_B, & 00|1_B0_A, \\
			& ~11|0_A0_B, & 11|0_B1_A, & 10|1_A1_B, & 11|1_B0_A\}\\
			\mathcal{E}_{BC}= &\{01|0_A0_C, & 00|0_C1_A, & 00|1_A1_C, & 00|1_C0_A, \\
			& ~10|0_A0_C, & 11|0_C1_A, & 11|1_A1_C, & 11|1_C0_A\}\\
			\mathcal{E}_{CA}= &\{00|0_C0_B, & 01|0_C1_A, & 00|1_A1_C, & 00|1_B0_C, \\
			& ~11|0_C0_B, & 10|0_C1_A, & 11|1_A1_C, & 11|1_B0_C\}\\
		\end{array}
	\end{equation}
	The graph of each of them  is a $4$-M\"{o}bius ladder $M_4$, we know that the classical bound, Tsirelson's bound  and exclusivity bound of three inequalities are $\alpha(M_4)=3$, $\vartheta(M_4)=2+\sqrt{2}$ and $\alpha^*(M_4)=4$ respectively. The power of EP shows up when we apply it to the overall exclusive set $\mathcal{E}=\mathcal{E}_{AB}\sqcup\mathcal{E}_{BC}\sqcup\mathcal{E}_{CA}$ with the corresponding testing parameter $\mathcal{I}=\mathcal{I}_{AB}+\mathcal{I}_{BC}+\mathcal{I}_{CA}$. There are some exclusive relations between the events of three different sets of events, for example $00|0_A0_B$ from $\mathcal{E}_{AB}$, $10|0_A0_C$ from $\mathcal{E}_{BC}$ and $11|0_C0_B$ from $\mathcal{E}_{CA}$ form a complete exclusive graph $K_3$. In fact, each column of the first three columns of the right-hand side of Eq. (\ref{3loop}) can be packed into two 3-complete graphs $K_3$, and the last column can be packed into $3$ 2-complete  graphs $K_2$, thus the graph $G_{\mathcal{E}}$ is packed into six $K_3$ graph and three $K_2$ graphs. The fractional packing number of the graph $G_{\mathcal{I}}$ is therefore $9$, i.e. $\mathcal{I}_{AB}+\mathcal{I}_{BC}+\mathcal{I}_{CA}\overset{E}{\leq} 9=R_C^{AB}+R_C^{BC}+R_C^{CA}$, this is exactly a monogamy relation.
	
	Note that the overall exclusive set is the disjoint union of three exclusive sets, this guarantees that the corresponding testing parameter $\mathcal{I}$ is exact the summation of three sub-testing parameters. This 3-loop nonlocality monogamy can also be derived from the non-signaling principle\cite{Jia2016}, Qin \emph{et. al.} provide a derivation completely from quantum mechanics\cite{Qin2015trade}. Here we prove that this phenomenon is a consequence of EP.
	
	As a generalization of this example, we apply Theorem 1 to Swetlichny's genuine multipartite nonlocal inequalities \cite{Svetlichny1987distinguishing,Collins2002bell,Seevinck2002bell,Bancal2011detecting}, the CHSH inequality is just a two partite special case of Svetlichny's inequality.
	
	\emph{Example~2.~Monogamy of Svetlichny's genuine nonlocality:}  Suppose $16$ experimenters are running three $4$-body genuine nonlocality test experiments $\mathcal{I}_4\leq 12$, see supplementary material \cite{supp} or Refs. \cite{Svetlichny1987distinguishing,Collins2002bell,Seevinck2002bell,Bancal2011detecting} for the explicit expression of the test parameter $\mathcal{I}_4$, two of them have the same expression ant the other is of the complementary form, then we have such a monogamy relation $\mathcal{I}_4^{ABCD}+\mathcal{I}_4^{CDEF}+\mathcal{I}_4^{EFAB}\overset{E}{\leq}36=R_C^{ABCD}+R_C^{CDEF}+R_C^{EFAB}$.
	
	Since we can packing the graph of $\mathcal{E}=\mathcal{E}_{ABCD}\sqcup \mathcal{E}_{CDEF}\sqcup \mathcal{E}_{EFAB}$ into 24 12-complete graphs $K_{12}$ and 12 8-complete graphs $K_8$, then $\alpha^*(G_{\mathcal{E}})=36= 3 R_C$, which is exactly a monogamy relation, see Supplemental Material \cite{supp} for more details. This is an example of a completely new type of monogamy relation, which is between genuine nonlocal correlations, the details are discussed in the appendix. Actually, in another work \cite{Jia2017genuine}, we give a more detailed investigation of such kind of monogamy.

	\emph{Example~3.~Monogamy of contextuality:} One of the initial successes of EP as a fundamental principle of quantum mechanics is that it can be used to derive the Tsirelson's bound of the  KCBS inequality. It is shown\cite{Ramanathan2012generalized,Jia2016} that two contextual correlations are monogamous. Here we give a very simple proof based on EP. As depicted in Fig. 2 (b), we have two KCBS inequalities $\mathcal{I}_{KCBS}$ and $\mathcal{I}'_{KCBS}$  each of which involves five dichotomic measurements $1,\cdots,5$ (respectively $1',\cdots,5'$), as in \cite{Ramanathan2012generalized}, we assume that the triple $1,1',2'$ are exclusive, \emph{viz.}, they can not get $0$ outcomes simultaneously, as is also the triple $4,5,5'$. Then we have two event sets $\mathcal{E}_{\mathcal{I}_{KCBS}}=\{01|12,01|23,01|34,01|45,01|51\}$ and $\mathcal{E}_{\mathcal{I}'_{KCBS}}=\{01|1'2',01|2'3',01|3'4',01|4'5',01|5'1'\}$, if we run two experiments simultaneously, we get a set $\mathcal{E}=\mathcal{E}_{\mathcal{I}_{KCBS}}\sqcup\mathcal{E}_{\mathcal{I}'_{KCBS}}$. By carefully analysis of the graph $G_{\mathcal{E}}$, we find that it can be packed into two $K_3$ graphs and two $K_2$ graphs, thus its packing number is  $\alpha^*(G_{\mathcal{E}})=4=R_C+R'_C$, this means that $\mathcal{I}_{KCBS}+\mathcal{I}'_{KCBS}\overset{E}{\leq} 4=R_C+R'_C$, we get the monogamy relation.
	
	Adopting proper exclusive assumptions, we can get many monogamy relations using theorem 1, like monogamy relations of n-cycle inequalities. See Supplemental Material\cite{supp} for more examples. All of these show that theorem 1 is a very general and useful result for monogamy relations, a large number of monogamy relations can be subsumed into this scenario.

	\emph{Exclusivity principle yields tight monogamy.}\textemdash Note that in all above examples we focus on the exclusivity relations appear in the isolated exclusivity set, \emph{viz.}, we have some restrictions $\sum_{e_i\in K_n} p(e_i)\leq 1$ for all $K_n\subseteq G_{\mathcal{E}}$. But EP can be applied to a broader scenario, which we refer to as the strong EP, it will give much more restrictions to calculate the exclusivity bound: $\sum p(e_i)p(\bar{e}_i)$, where $\bar{e}_i$ is corresponding events of $e_i$ in the complementary graph of $G_{\mathcal{E}}$.
	Actually, in many cases, it will make the exclusivity bound equal to the quantum bound, this is exactly the meaning that EP tight bound the quantum correlations\cite{Yan2013}. we now analyze how this principle can be applied to the monogamy phenomenon to give the tight monogamy bound restricted by quantum mechanics. If two experiments are implemented simultaneously, their monogamy score cannot always meet the bound $R_{C}^1+R_{C}^2$. With the strong EP, we have an explicit bound of the monogamy score restricted by quantum mechanics:
	
	\emph{Theorem~2.} Let $\mathcal{E}_1,\cdots, \mathcal{E}_n$ be several disjoint experimental event sets, the integral event set $\mathcal{E}=\sqcup_{i=1}^n\mathcal{E}_i$ is the disjoint union of these sets. Their monogamy score is given by the Lov\'{a}sz number $\vartheta(G_{\mathcal{E}})$ of the integral exclusivity graph $G_{\mathcal{E}}$. These experiments are monogamous if and only if
	\begin{equation}\label{}
		\vartheta(G_{\mathcal{E}})\leq \sum_i \alpha (G_{\mathcal{E}_i}),
	\end{equation}
	the monogamy is tight if equality holds.
	
	See supplementary material for the proof. Note that Theorem 1 is an operational weak version of this theorem since Lov\'{a}sz number of a graph is more difficult to calculate than the fractional packing number of a graph. If some experiments satisfy the conditions of theorem 1, they of course satisfy the condition of theorem 2, \emph{viz.}, and they are monogamous. Theorem 1 is the monogamy relations for any generalized probability theory which obeys EP and theorem 2 is the quantum version.
	
	\emph{Discussions and conclusions}\textemdash In this letter we investigated the monogamy of correlation inequality and indicated that the origin of the phenomenon is the EP. We gave the necessary and sufficient conditions for several correlation experiments to be monogamous. Besides, we give some new types of monogamy relations, in particular, we give the monogamy of genuine nonlocality, its origin is still exclusivity. Our work provides some new evidence for EP being a basic physical principle to give the prediction of quantum mechanics.

	\indent Z.-A. J. acknowledges  Bai-Chu Yu, Dong-Hui Yang, Biao Yi, Sheng Tan and Shi-Yu Cao for many beneficial discussions. This work is supported by the National Key R \& D Program (Grant No. 2016YFA0301700), the Strategic Priority Research Program of the Chinese Academy of Sciences (Grants No. XDB01030100 and No. XDB01030300), and the National Natural Science Foundation of China (Grants No. 11275182 and No. 61435011).

	\onecolumngrid
	\clearpage

	\begin{center}
		\large \bf Supplemental Material
	\end{center}

	\twocolumngrid

	\appendix

	\section{Graph theoretical terminologies}
	
	A graph $G=(V,E)$ is a pair of sets such that $E\subseteq [V]^2$, where $[V]^2$ denotes the family of 2-element subsets of $V$. We call the set $V=V(G)$ the vertex set of $G$ and $E=E(G)$ the edge set of $G$. Two vertices $v_i$ and $v_j$ are adjacent if there exist an edge $e\in E$ such that $v_{i} \perp v_{j}\in e$. A vertex weight is a map $w:V(G)\to \mathbb{R}$ and a weighted graph is a graph with a vertex weight.

	The independent number $\alpha(G)$ of the graph $G$ is the cardinality of the maximum independent vertex set, where independent vertex means that each pair of vertices in the set are not adjacent.
	
	The orthonormal representation a a graph $G$ is a map $r$ from vertex set $V(G)$ of G to some vector space $W$, for which each vector $r(v_i)$ is a unit vector and $r(v_i)$ and $r(v_j)$ are orthogonal if $v_i$ and $v_j$ are adjacent. For simplicity, we will use the same letter $v_i$ to label the vector corresponding to vertex $v_i$.  The Lov\'{a}sz number $\vartheta(G)$ is then defined as
	\begin{equation}\label{}
		\vartheta(G)=\max \sum_{i} |\langle v_i|\phi\rangle|^2,
	\end{equation}
	where the maximum is taken over all orthonormal representations, i.e., in all dimensions and all orthonormal vector assignments of that dimension, and over all unit vector $\phi$.
	
	The fractional packing number of the graph $G$ is the maximal value $\sum_{v_i\in V(G)}p(v_i)$ with the constraints $\sum_{v\in K_n\subseteq G} p(v)\leq 1$ for all complete subgraphs $K_n$ of $G$, where $p(v)$ is non-negative real number less than one.
	
	Here we list some important properties of independent number $\alpha (G)$, Lov\'{a}sz number $\vartheta(G)$ and  fractional packing number $\alpha^*(G)$:
	
	\begin{itemize}
		\item For any graph $G$, we have $\alpha(G)\leq \vartheta(G)\leq \alpha^*(G)$;
		\item For the OR product graph $G*H$, $\vartheta(G*H)=\vartheta(G)\vartheta(H)$;
		\item For the odd cycle  graph $C_n$, $\vartheta(C_n)=\frac{n\cos (\pi/n)}{1+\cos(\pi/n)}$.
	\end{itemize}

	\section{Proof of theorem 1 and theorem 2 in the main text}
	\emph{Theorem~1.} Given several disjoint experimental event sets $\mathcal{E}_1,\cdots, \mathcal{E}_n$ with exclusive graphs $G_{\mathcal{E}_1},\cdots,G_{\mathcal{E}_n}$ respectively, we can make them into an assemblage of events $\mathcal{E}=\sqcup_{i=1}^n\mathcal{E}_i$ with exclusive graph $G_{\mathcal{E}}$, if we have the relation
	\begin{equation}\label{}
		\alpha^*(G_{\mathcal{E}})\leq \sum_{i=1}^n \alpha(G_{\mathcal{E}_i}),
	\end{equation}
	then these experiments are monogamous.
	
	\emph{Proof.} In fact, we need to translate graph terms into their physical correspondences: $\mathcal{E}=\sqcup_{i=1}^n\mathcal{E}_i\leftrightarrow \mathcal{I}_{\mathcal{E}}=\sum_i\mathcal{I}_{\mathcal{E}_i}$, $\alpha^*(G_{\mathcal{E}})\leftrightarrow R_E(\mathcal{E})$ and $\alpha(G_{\mathcal{E}_i})\leftrightarrow R_C(\mathcal{E}_i)$. Thus we arrive at the monogamy relation: $\sum_{i}\mathcal{I}_{\mathcal{E}_i}\overset{Q}{\leq}R_Q(\mathcal{E})\overset{E}{\leq}R_E(\mathcal{E})\leq \sum_i R_C(\mathcal{E}_i)$.

	\emph{Theorem~2.} Let $\mathcal{E}_1,\cdots, \mathcal{E}_n$ be several disjoint experimental event sets, the integral event set $\mathcal{E}=\sqcup_{i=1}^n\mathcal{E}_i$ is the disjoint union of these sets. Their monogamy score is given by the Lov\'{a}sz number $\vartheta(G_{\mathcal{E}})$ of the integral exclusivity graph $G_{\mathcal{E}}$. These experiments are monogamous if and only if
	\begin{equation}\label{}
		\vartheta(G_{\mathcal{E}})\leq \sum_i \alpha (G_{\mathcal{E}_i}),
	\end{equation}
	the monogamy is tight if equality holds.
	
	\emph{Proof.} Actually we can regard these $n$ experiments as a whole experiment $\mathcal{E}$, the quantum bound $R_Q(\mathcal{E})$ of this experiment is $\vartheta(G_{\mathcal{E}})$, thus we arrive at $\mathcal{I}=\sum_{i=1}^n\mathcal{I}(\mathcal{E}_i)\overset{Q}{\leq}R_Q=\vartheta(G_{\mathcal{E}})\leq \sum_{i=1}^n \alpha (\mathcal{E}_i)=\sum_{i=1}^n R_C(\mathcal{E}_i)$, which is the monogamy relation.
	
	\section{ Monogamy of Genuine Nonlocality}
	Compared with the two-body nonlocality, the $n$-body nonlocal correlations have a richer structures. Svetlichny\cite{Svetlichny1987distinguishing} argued that there may exist some tripartite correlations which can not be described using the so-called hybrid LHV models:
	\begin{align}\label{}
		p(a,b,c)= & P_{A|BC}\int d\lambda p(a|\lambda)p(b,c|\lambda)\nonumber \\
		& +P_{C|AB}\int d\lambda p(c|\lambda)p(a,b|\lambda)\nonumber\\
		&+P_{B|AC}\int d\lambda p(b|\lambda)p(a,c|\lambda),
	\end{align}
	where $P_{A|BC}+P_{C|AB}+P_{B|AC}=1$. He provided an inequality to test this kind of nonlocality and find that some quantum states can violate the inequality. This inequality was later generalized into the $n$-body case\cite{Collins2002bell,Seevinck2002bell,Bancal2011detecting}, which is of the form
	\begin{equation}\label{}
		S_n\overset{hLVV}{\leq} 2^{n-1}\overset{Q}{\leq} \sqrt{2}\times 2^{n-1},
	\end{equation}
	where $S_n$ is defined recursively from CHSH parameter $S_2=0_10_2+0_11_2+1_10_2-1_11_2$, the subscripts are used to denote different party. The recursive formula is
	\begin{equation}\label{}
		S_n=S_{n-1}1_{n}+\bar{S}_{n-1}0_n,
	\end{equation}
	where $\bar{S}_{n-1}$ is obtained from $S_{n-1}$ by exchanging $0_i$ and $1_i$ for all $i=1,\cdots, n-1$. The Svetlichny inequality can be rewritten as a linear inequality of probabilities of involved events as
	\begin{equation}\label{}
		\mathcal{I}_n\overset{C}{\leq} 3\times 2^{n-2}\overset{Q}{\leq} (2+\sqrt{2})\times 2^{n-2},
	\end{equation}
	where $\mathcal{I}_n$ takes the form
	\begin{eqnarray}\label{}
		\begin{array}{c}
			\sum_{\begin{subarray}{lcl}
					(A_1,A_2,A_3)\neq (A_1,A_1,A_1)\\a_1\oplus \cdots \oplus a_n=0\end{subarray}}p(a_1,\cdots, a_n|A_1,\cdots, A_n)\\
			+\sum_{\begin{subarray}{lcl}
					(A_1,A_2,A_3)= (A_1,A_1,A_1)\\a_1\oplus \cdots \oplus a_n=1\end{subarray}}p(a_1,\cdots, a_n|A_1,\cdots, A_n)
		\end{array}
	\end{eqnarray}
	
	The work of Cabello\cite{Cabello2015simple} indicates that the quantum bound of $S_n$ can be derived from the exclusivity principle.
	
	Here we derive the monogamy of three $S_4$ from the exclusivity principle using theorem 1. The precise formula of $S_4$ is
	\begin{equation}\label{}
		\begin{array}{cl}
			S_4 =&|\langle -1111+1101+0111+1011 \\
			&-0000+0010+1000+0100 \\
			&-0001+0011+1001+0101 \\
			&-1110+1100+0110+1010 \rangle|.
		\end{array}
	\end{equation}
	It can be equivalently expressed as a linear inequality of probability of events as
	\begin{equation}\label{}
		\begin{array}{cc}
			\mathcal{I}_4 = & \sum_{a\oplus b\oplus c \oplus d =0} p(abcd|1111)\\
			&+\sum_{a\oplus b\oplus c \oplus d =1} p(abcd|1101)\\
			&+\sum_{a\oplus b\oplus c \oplus d =0} p(abcd|0111)\\
			&+\sum_{a\oplus b\oplus c \oplus d =0} p(abcd|1011)\\
			&+\sum_{a\oplus b\oplus c \oplus d =1} p(abcd|0000)\\
			&+\sum_{a\oplus b\oplus c \oplus d =0} p(abcd|0010)\\
			&+\sum_{a\oplus b\oplus c \oplus d =0} p(abcd|1000)\\
			&+\sum_{a\oplus b\oplus c \oplus d =0} p(abcd|0100)\\
			&+\sum_{a\oplus b\oplus c \oplus d =1} p(abcd|0001)\\
			&+\sum_{a\oplus b\oplus c \oplus d =0} p(abcd|0011)\\
			&+\sum_{a\oplus b\oplus c \oplus d =0} p(abcd|1001)\\
			&+\sum_{a\oplus b\oplus c \oplus d =0} p(abcd|0101)\\
			&+\sum_{a\oplus b\oplus c \oplus d =1} p(abcd|1110)\\
			&+\sum_{a\oplus b\oplus c \oplus d =0} p(abcd|1100)\\
			&+\sum_{a\oplus b\oplus c \oplus d =0} p(abcd|0110)\\
			&+\sum_{a\oplus b\oplus c \oplus d =0} p(abcd|1010).
		\end{array}
	\end{equation}
	
	Suppose twelve physicists $A, \cdots, F$ run three genuine nonlocality test experiments simultaneously, with $\mathcal{I}_4^{ABCD}$ and $\mathcal{I}_4^{CDEF}$ of the same form whilst $\mathcal{I}_4^{EFAB}$ are of the complementary form. Their exclusive event sets are $\mathcal{E}_{ABCD}$, $\mathcal{E}_{CDEF}$ and $\mathcal{E}_{EFAB}$ respectively. The integral event set is $\mathcal{E}=\mathcal{E}_{ABCD}\sqcup \mathcal{E}_{CDEF}\sqcup \mathcal{E}_{EFAB}$. We will show that the graph $G_{\mathcal{E}}$ corresponds to the integral event set that can be packed into 24 12-complete graphs $K_{12}$ and 12 8-complete graphs $K_8$. Thus $\alpha^*(G_{\mathcal{E}})=36=3\times R_C$ which means that $\mathcal{I}_4^{ABCD}+\mathcal{I}_4^{CDEF}+\mathcal{I}_4^{EFAB}\overset{Q}{\leq} R \overset{E}{\leq} 3R_C$, they can not simultaneously be violated.
	
	The events in the experiment set $\mathcal{E}$ are listed in Table \ref{tab:table1}, where `$+$' means that the module 2 sum of all outcomes is equal to 0, and `$-$' for 1.
	\begin{table}
		\caption{\label{tab:table1}In this table we list the packing set of the whole experiment set.}
		\begin{ruledtabular}
			\begin{tabular}{ccc}
				$S^{ABCD}_4$ &$S^{CDEF}_4$ &$S^{EFAB}_4$\\
				\hline
				$-abcd|1111$ & $-cdef|1111$ & $+efab|1111$\\
				$+abcd|1100$ & $+cdef|0010$ & $-efab|1011$\\
				$+abcd|1101$ & $+cdef|0101$ & $-efab|0111$\\
				$-abcd|1110$ & $-cdef|1000$ & $-efab|0011$\\
				
				$-abcd|0000$ & $-cdef|0000$ & $+efab|0000$\\
				$-abcd|0001$ & $-cdef|0111$ & $-efab|1100$\\
				$+abcd|0010$ & $+cdef|1010$ & $-efab|1000$\\
				$+abcd|0011$ & $+cdef|1101$ & $-efab|0100$\\
				
				$+abcd|0111$ & $+cdef|1110$ & $-efab|1001$\\
				$+abcd|0100$ & $+cdef|0001$ & $-efab|0101$\\
				$+abcd|0101$ & $+cdef|0100$ & $+efab|0001$\\
				$+abcd|0110$ & $+cdef|1011$ & $-efab|1101$\\
				
				$+abcd|1011$ & $+cdef|1100$ & $-efab|0010$\\
				$+abcd|1000$ & $+cdef|0011$ & $+efab|1110$\\
				$+abcd|1010$ & $+cdef|1001$ & $-efab|0110$\\
				$+abcd|1001$ & $+cdef|0110$ & $-efab|1010$\\
			\end{tabular}
		\end{ruledtabular}
	\end{table}
	
	Each column with an odd number of `$-$' can be packed into two $12$-complete graph $K_{12}$, we take the second row as an example, it can be divided as $\mathcal{E}^{(1)}$ of the form
	\begin{equation*}
		\left\{
		\begin{array}{ccccc}
			ABCD& 0000|1100 & 0011|1100 & 1100|1100 & 1111|1100 \\
			CDEF& 0101|0010 & 0110|0010 & 1001|0010 & 1010|0010 \\
			EFAB& 1101|1011 & 1110|1011 & 0001|1011 & 0010|1011
		\end{array}\right\},
	\end{equation*}
	and $\mathcal{E}^{(2)}$ of the form
	\begin{equation*}
		\left\{
		\begin{array}{ccccc}
			ABCD& 0101|1100 & 0110|1100 & 1001|1100 & 1010|1100 \\
			CDEF& 1100|0010 & 1111|0010 & 0000|0010 & 0011|0010 \\
			EFAB& 0111|1011 & 0100|1011 & 1011|1011 & 1000|1011
		\end{array}\right\}.
	\end{equation*}
	Similarly, we can pack another row into two $K_{12}$ graphs. There are a totally 24 $K_{12}$ graphs.
	
	For the row with an even number of `$-$', we can pack it into their $K_8$ graphs, this is obvious since for each fixed measurement, there are eight outcomes, and there events are pairwise exclusive. Thus row 1, 5, 11, and 14 can be packed into 12 $K_8$ graphs. This completes the proof. In another work Jia \emph{et. al.} give a more complete investigation of Svetlichny's genuine nonlocality, see Ref.\cite{Jia2017genuine} for detail.

	\section{More Examples of monogamy}
	In this section, we give the monogamy relations between $n$-cycle non-contextual inequalities, here we derive the monogamy relations from the exclusivity principle, which is different from the approach in Ref.\cite{Jia2016}.
	
\begin{figure}
		\includegraphics[scale=1.0]{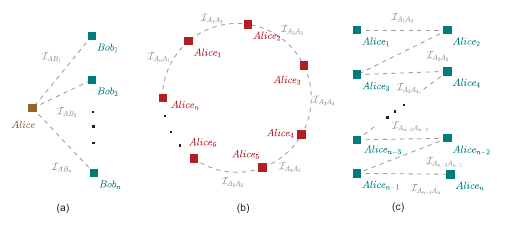}
		\caption{\label{fig:result} (color online). The depiction of the monogamy relations, where each vertex represents an experimenter and two vertices are adjacent if they run an experiment.}
	\end{figure}

	(i)Suppose that $n$ Alice do $2m$-cycle correlation test experiments $\mathcal{I}_{A_1A_2},\cdots ,\mathcal{I}_{A_nA_1}$ as depicted in Fig.~\ref{fig:result}(b), and the $i$-th Alice choose $m$ observables $1_{i},\cdots, m_{i}$. If each pair of $k$-th obervables $k_{i}$ and $k_{j}$ of two nonadjacent Alice(i.e., $i\neq j\pm 1$) is a xor pair(i.e. events $a|k_{i}$ and $b|k_{j}$ are exclusive for $a=b$ ), then we have a loop-type monogamy relation:
	\begin{equation}\label{}
		\mathcal{I}_{A_1A_2}+\cdots+\mathcal{I}_{A_nA_1}\overset{E}{\leq} R_E\leq R_{C}^{12}+\cdots+R_{C}^{n1}.
	\end{equation}
	
	(ii) In the one-to-many scenario, Alice runs $n$ experiments with $n$ Bob, Alice's observables are chosen from $\{1_A,\cdots,m_A\}$ and the $j$-th Bob's observable set is $\{k_j\}_{k=1}^m$, again we assume that each pair of $k$-th observables $k_{i}$ and $k_{j}$ of two different Bob is a xor pair, then there is a monogamy relation like:
	\begin{equation}
		\mathcal{I}_{AB_1}+\cdots+\mathcal{I}_{AB_n}\overset{E}{\leq} R_E \leq R_{C}^1+ \cdots+R_{C}^n
	\end{equation}
	
	(iii) As depicted in Fig \ref{fig:result}(c), we also have a chain type monogamy relation
	\begin{equation}\label{}
		\mathcal{I}_{A_1A_2}+\cdots+\mathcal{I}_{A_{n-1}A_n}\overset{E}{\leq} R_E \leq R_{C}^{12}+\cdots+R_{C}^{n-1n}.
	\end{equation}
	\emph{Proof.} (i) The event set of each experiment $\mathcal{I}_{A_iA_{i+1}}$ is
	\[
	\begin{array}{ccccc}
		\mathcal{E}_{i,i+1}= &\{00|1_i1_{i+1}, & 00|1_{i+1}2_i, &\cdots, & 01|m_{i+1}1_i, \\
		& ~11|1_i1_{i+1}, & 11|1_{i+1}2_i, & \cdots, & 10|m_{i+1}1_i\}.
	\end{array}
	\]
	We need to calculate the packing number $\alpha^*(G_{\mathcal{E}})$ for the overall event set $\mathcal{E}=\sqcup_{i=1}^n \mathcal{E}_{i,i+1}$. Note that we can divide $\mathcal{E}$ into $2m$ groups as
	\[
	\begin{array}{ccccc}
		\mathcal{E}^1= &\{00|1_11_{2},&00|1_21_{3}, & \cdots, & 00|1_n1_{1},\\
		&~11|1_11_{2},&11|1_21_{3}, & \cdots, & 11|1_n1_{1}\}\\
		\vdots &  &\vdots &  &\\
		\mathcal{E}^{2m}= &\{01|m_21_{1},& 01|m_31_{2}, & \cdots, & 01|m_11_n,\\
		&~10|m_21_{1},& 10|m_31_{2}, & \cdots, & 10|m_1i_n\}.
	\end{array}
	\]
	Since the packing number of $\mathcal{E}^j(j\leq 2m-1)$ is $\alpha(G_{\mathcal{E}^j})=\frac{2n}{3}$ and for $\mathcal{E}^{2m}$ it is $\alpha^*(G_{\mathcal{E}^{2m}})=n$, thus $\alpha^*(G_{\mathcal{E}})\leq (2m-1)\frac{2n}{3}+n$. But note that the sum of all classical bound of these testing inequalities satisfies $\sum R_C^{ii+1}=n(2m-1)\leq (2m-1)\frac{2n}{3}+n$ for $m\geq 2$. Therefore,
	\[
	\begin{array}{ll}
		\mathcal{I_{\mathcal{E}}}&=\sum_{i=1}^n \mathcal{I}_{A_iA_{i+1}}\overset{E}{\leq} \alpha^*(G_{\mathcal{E}})\leq (2m-1)\frac{2n}{3}+n \\
		& \leq \sum_{i=1}^n R_C^{ii+1},
	\end{array}
	\]
	this completes the proof.
	
	In the same spirit as the proof of (i), we collect all events and redivide them into some disjoint subsets, we find that the packing number of the overall event set $\mathcal{E}$ is less than the sum of the classical bound of each inequality: $\alpha^*(G_{\mathcal{E}})\leq \sum R_{C}$. We can prove (ii) and (iii).\qed

	Here, we use the term Alice and Bob only for the convenience of description, they are not assumed to be spatially separated like in Bell's scenario. To see how the conditions for nonadjacent Alice's observables can be implemented in quantum mechanics, let us take the $i$-th Alice's observable set as $\{k_{i}=2|v_i^k\rangle \langle v_i^k|-\mathds{1}\}$ and the $j$-th Alice's observable set as $\{k_{j}=2|v_j^k\rangle \langle v_j^k|-\mathds{1}\}$ with $\langle v_i^k|v_j^k\rangle=0$ for all $k$ and $i\neq j\pm 1$. It is obvious that $k_{i}$ and $k_{j}$ are commutative and they cannot simultaneously have the same outcomes. This implementation is a much stronger condition than the one in the derivation of monogamy from the no-disturbance principle where $k_{i}$ and $k_{j}$ are merely commutative\cite{Jia2016}.
	
	\bibliographystyle{apsrev4-1-title}

\end{document}